\documentclass[pra,twocolumn,showpacs,superscriptaddress,amsmath,amssymb]{revtex4}

\usepackage{graphicx}

\begin{document}


\title{Resonant trapping in the transport of a matter-wave soliton through a quantum well}

\author{Thomas Ernst}
\affiliation{%
Centre for Theoretical Chemistry and Physics and Institute for Natural Sciences, Massey University (Albany campus), Private Bag 102 904, NSMC, Auckland, New Zealand
}%
\author{Joachim Brand}
\affiliation{%
Centre for Theoretical Chemistry and Physics and Institute for Natural Sciences, Massey University (Albany campus), Private Bag 102 904, NSMC, Auckland, New Zealand
}%


\date{\today}

\begin{abstract}
We theoretically investigate the scattering of bright solitons in a Bose-Einstein condensate on narrow attractive potential wells. 
Reflection, transmission and trapping of an incident soliton are predicted to occur with remarkably abrupt transitions upon varying the potential depth.
Numerical simulations of the nonlinear Schr{\"o}dinger equation are complemented by a variational collective coordinate approach.
The mechanism for nonlinear trapping is found to rely both on resonant interaction between the soliton and bound states in the potential well as well as radiation of small amplitude waves. These results suggest that solitons can be used to probe bound states that are not accessible through scattering with single atoms.
\end{abstract}

\maketitle

\section{\label{intro}Introduction}

A classical particle incident on a potential barrier can either reflect or pass over it, depending on its kinetic energy in relation to the height of the barrier. Therefore it always passes a negative barrier, e.g. a hole. This is not the case any more for quantum mechanical particles or matter waves \cite{friedrich2004www}. Solving the linear Schr\"odinger equation shows that partial reflection can be expected from attractive potentials such as quantum wells. In recent experiments by Pasquini et al. \cite{2004PhRvL..93v3201P,2006PhRvL..97i3201P} Bose-Einstein condensates (BECs) have been found to reflect from a  surface in spite of mostly attractive atom-surface interactions. In addition to the wave nature of ultra-cold atoms, BECs also experience nonlinear mean-field interactions, which present a complication in the experiments \cite{scott2005aqr,scott2007nds}. On the other hand, the nonlinear interactions are potentially useful when they result in effects that are not obtainable with linear matter waves.

Here we focus on BECs with attractive interactions confined to a quasi-one-dimensional waveguide that can form self-localised wave packets known as bright solitons. In the limit of tight transverse confinement, the mean-field theory of BECs reduces to the nonlinear Schr\"odinger equation, an integrable soliton equation \cite{pitaevskiui2003bec}. A well-known property of solitons is that they behave in many respects as classical particles do, e.g.\ with respect to their collisional properties or their motion in a slowly-varying external potential \cite{1989RvMP...61..763K}. As solitons have both particle and wave properties, they may experience ``quantum'' reflection from an attractive potential well and yet maintain their particle-like integrity to a large degree \cite{2006EL.....73..321L}. In addition to such nonlinear wave effects, macroscopic quantum tunneling and fragmentation have recently been discussed \cite{weiss:010403,Streltsov2009}.

On the other hand there is a possibility that a scattering soliton, or part of it, may become trapped in a potential well. Such an effect would not be possible for either a classical or a quantum mechanical particle in the absence of dissipation. 
In this paper we study the effect of trapping of solitons in attractive potential wells. We argue that a resonant population transfer between the soliton and non-linear bound states of the potential well, first suggested by Goodman, Holmes, and Weinstein \cite{2004PhyD..192..215G}, are vital for this effect. Here, we explore the details and the consequences of this mechanism. We find that it is possible to deduce the energy of the most weakly bound state of an unknown defect by scattering solitons with known parameters and by recording the trapped particle number. This could potentially lead to real-world applications of nonlinear-wave scattering. Although the current work addresses matter-wave solitons in particular, our findings are equally applicable to nonlinear optics or other nonlinear wave problems governed by the nonlinear Schr\"odinger equation.

Recent progress in experiments has made it possible to create matter-wave solitons and to explore their properties \cite{2002Sci...296.1290K,2002Natur.417..150S,PhysRevLett.92.230401}. This, besides the importance of soliton physics in other areas of physics \cite{zabusky1965isc,rajaraman1982sai,hasegawa1973tsn,kivshar2003os} has motivated a variety of authors to investigate the scattering of solitons on different kinds of potentials, like barriers or impurities \cite{forinash1994idb,cao1995sdc,frantzeskakis2002ids,2003Chaos..13..874M,2007NJPh....9....4A,weiss:010403,Streltsov2009}, wells \cite{2004JPSJ...73..503S,2004PhyD..192..215G,2004PhRvE..70f6622S,2006OptL...31..966M,2006EL.....73..321L}, steps \cite{aceves1989tlb,kivshar1990ret,Cornish20091299} and a potential ramp \cite{1996PhRvE..53.2823F}. Here, we focus on the quantum well, as which we understand an attractive potential well with well distinguished single-particle energy levels representing linear bound states.
One of us has previously investigated the enhancement of quantum reflection by nonlinear interactions in solitons and the abrupt transition to transmission in the scattering on an attractive defect potential \cite{2006EL.....73..321L}. The aim of the current work is to extend the previous work to include trapping phenomena. Varying the strength of the attractive defect and scattering a slow soliton, we identify regimes dominated by transmission, reflection, trapping, and a combination of trapping and reflection.
We discuss a trapping mechanism by a resonant transfer to (quasi-) bound states within a quantum 
well. Similar resonant effects have been investigated in the transport of repulsive condensates through a double barrier potential \cite{2005PhRvL..94b0404P}.

In Sec. \ref{sec:level0} of this paper we introduce the theoretical approach employed for scattering matter-wave bright solitons in a tight waveguide trap.
Numerical simulations of trapping phenomena are presented in Sec. \ref{sec:level1} before discussing resonant mechanisms for trapping and transmission. A collective coordinate approach based on a variational model is discussed in  Sec.\ \ref{sec:level2} and compared with the simulation results.
Sec.\ \ref{sec:level3} then discusses the trapping mechanism in more detail. The final section Sec.\ \ref{sec:level4} discusses how energy levels of defects can be probed via scattering of solitons.

\section{\label{sec:level0}Theoretical model} 

We consider an attractively interacting BEC in a waveguide-like trap with tight harmonic confinement in two dimensions but weak or no confinement in the remaining ($z$) dimension. In such a situation, bright solitons constituting localized (bound) BEC wave packets are metastable and collapse ensues beyond a critical particle number \cite{carr2002dmw}. We assume that the soliton size remains sub-critical and that the linear density $n(z,\tau)$ at any time remains well below the threshold for transverse collapse $- n a_s \ll 11.7/(8\pi)$ \cite{weinstein1983nse,PhysRevLett.92.040401}, where $a_s$ is the $s$-wave scattering length. In this case, the soliton dynamics may be modeled with the one-dimensional Gross-Pitaevskii (GP) equation
\begin{equation} \label{eq:1DGPE}
  i\hbar\frac{\partial}{\partial \tau} \phi= \left[ -\frac{\hbar^2}{2m}\frac{\partial^2}{\partial z^2} + g_{1D} |\phi|^2 + V_{1D}(z)\right]  \phi ,
\end{equation}
where the GP wave function $|\phi(z,\tau)| = \sqrt{n(z,\tau)}$ is normalized to the number of atoms $\int |\phi|^2 dz = N$ and the 1D  interaction constant is $g_{1D}\approx2\hbar\omega_\bot a_s$ \cite{PhysRevLett.81.938,PhysRevA.70.043622} under the influence of a transverse harmonic trapping potential with frequency $\omega_\bot$. The quasi-one-dimensional approximation (\ref{eq:1DGPE}) is expected to break down for solitons of close to critical size, where the dynamics may become inherently three-dimensional \cite{parker2008cbs}. Small quantitative corrections to Eq.~(\ref{eq:1DGPE}) due to the finite transverse extent of the solitons \cite{PhysRevA.66.043603,sinha2006fad} are not expected to significantly alter the results reported below and are thus neglected.

Choosing an energy scale $\tilde{E} > 0$ and a density scale $\tilde{n}$ 
we can rewrite Eq.~(\ref{eq:1DGPE}) in dimensionless form by introducing $t=\tau/\tilde{t}$, $x=z/\tilde{x}$, $\psi(x)=\phi(z)/\sqrt{\tilde{n}}$, where $\tilde{t} = \hbar/\tilde{E}$ and $\tilde{E} = \hbar^2/(m \tilde{x}^2)$:
\begin{equation}
 i\frac{\partial}{\partial t}\psi (x,t)=\left[ -\frac{1}{2}\frac{\partial^2}{\partial x^2} +g|\psi (x,t)|^2 +V(x)\right] \psi (x,t) .
\label{GPE-soliton}
\end{equation}
where we have introduced $V(x)=V_{1D}(x)/\tilde{E}$ and the dimensionless coupling constant $g = g_{1D} {\tilde{n}}/\tilde{E}$. In the following we assume $g<0$ since we consider attractive BECs that support bright solitons.
At this point the energy $\tilde{E}$ and  density scale $\tilde{n}$ remain arbitrary and can be chosen to suit experimental parameters. We will discuss specific choices below.

For a vanishing potential $V(x)$ Eq.~(\ref{GPE-soliton}) has the soliton solution
\begin{equation}
 \psi(x)=A\mathrm{sech}\left(A\sqrt{-g}(x-x_0-vt)\right) \exp(i\theta (x,t)) ,
\label{soliton_solution}
\end{equation}
where $v$ is the velocity of the soliton measured in units of $\tilde{x}/\tilde{t}$ and $x_0$ is the dimensionless position at $t=0$.
The solution is normalized according to
\begin{equation}
 {\cal N}_s = \int dx |\psi (x,t)|^2 =\frac{2A}{\sqrt{-g}} ,
 \label{wf-normalization}
\end{equation}
where $A$ is a dimensionless amplitude and ${\cal N}_s$ is related to the particle number by $N = {\cal N}_s \tilde{n} \tilde{x}$.
The phase is given by $\theta (x,t)=vx-\omega t$ and  $\omega = {v^2}/{2}+ \mu$
is the
dimensionless frequency. Here, $\mu = {g} A^2 /2$ is the (negative) chemical potential of a stationary soliton measured in units of $\tilde{E}$ .
The soliton width in units of $\tilde{x}$ is given by $l_s=1/(A\sqrt{-g})$ \cite{1989RvMP...61..763K}.

For the numerical simulations below we will commonly choose
$g=-1$, which relates the energy and density scales by $\tilde{E} = g_{1D} \tilde{n}$. The further choice of $A=1$ fully determines the energy scale $\tilde{E} = m N^2 g_{1D}^2 /(4 \hbar^2) = N^2 \omega_\perp^2 a_s^2 m$ and the density scale $\tilde{n} = N/(2 \tilde{x})$, where the unit length becomes  $\tilde{x} = -\hbar / (N \omega_\perp a_s m)$ and the dimensionless soliton length becomes $l_s=1$.

Typical experimental values for $^7\mathrm{Li}$ BEC \cite{2002Natur.417..150S,2002Sci...296.1290K} are $\omega_\bot \approx 2\pi \times 710\mathrm{Hz}$, $a_s\approx -0.2\mathrm{nm}$ and $N\approx 6 \times 10^3$. 
This yields a length scale of $\tilde{x} \cong 1.7\mu$m, which is consistent with experimental observations \cite{2002Sci...296.1290K}.
For these parameters the time unit is $\tilde{t} \approx0.3$ms.  The velocity scale is consequently $\tilde{x}/\tilde{t} \approx 5.7\mu$m/ms.
\begin{figure}
\includegraphics[width=0.7\columnwidth,angle=-90]{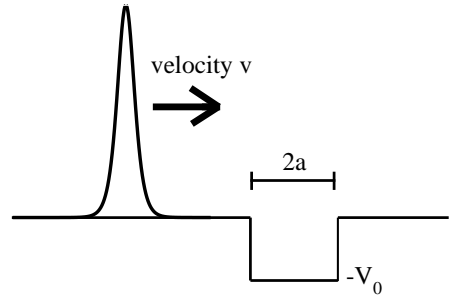}
\caption{\label{fig:moving-soliton} A soliton is being scattered on a rectangular well with width $2a$ and depth $V_0$.}
\end{figure}

For the form of the external potential we consider a rectangular well defined as
\begin{equation}
V(x) = \left\{
 \begin{array}{ccc}
 0 & \mathrm{for} & |x|>a \\
-V_0 & \mathrm{for} & |x|\leq a.
\end{array}
\right.
\label{eq:well}
\end{equation}
as shown in Fig.\ \ref{fig:moving-soliton}. We are specifically interested in the case where the width of the well $2a$ is comparable to the soliton width $l_s$ and therefore we choose $2a=1$ for the numerical studies in this work unless noted otherwise.

Lee and Brand \cite{2006EL.....73..321L} have already investigated in detail the enhanced reflection of solitons for the special case of a Rosen-Morse potential $-V_0\mathrm{sech}^2(\alpha x)$  at low velocities. There they found a step-like behavior of the reflection and transmission probabilities, which is due to the nonlinearity. Here, we aim to extend this work to include resonant trapping effects.

We solved Eq.~(\ref{GPE-soliton}) numerically via the Crank-Nicholson method using a standard finite difference discretization of the spatial derivatives \cite{pressNumericalRecipes}. The algorithms were implemented in standard C/C++ and Octave \cite{eaton:2002}. 
We performed the simulations in a box with hard wall boundaries. The box length was set to $l_{box}=80l_s$ unless stated otherwise. Furthermore we used $N_g=2001$ grid points and a fixed time step of $\Delta t=0.01$.
The convergence of our calculations with respect to these quantities was monitored carefully. Reflection from the boundaries was avoided by appropriate timing the the simulation. We also used complex absorbing potentials at the boundaries for verifying that reflection effects remained below a quantifiable threshold.

\section{\label{sec:level1}Phenomenology of soliton scattering on a quantum well}

In this section we present results from numerical solutions of Eq.~(\ref{GPE-soliton}) corresponding to a soliton approaching the well of Eq.~(\ref{eq:well}).
In the initial setup the soliton  (\ref{soliton_solution}) is being placed at position $x_0 = -12$ left  of the quantum well moving with the velocity $v>0$ towards it (see Fig. \ref{fig:moving-soliton}).

\begin{figure}
\includegraphics[width=0.7\columnwidth,angle=-90]{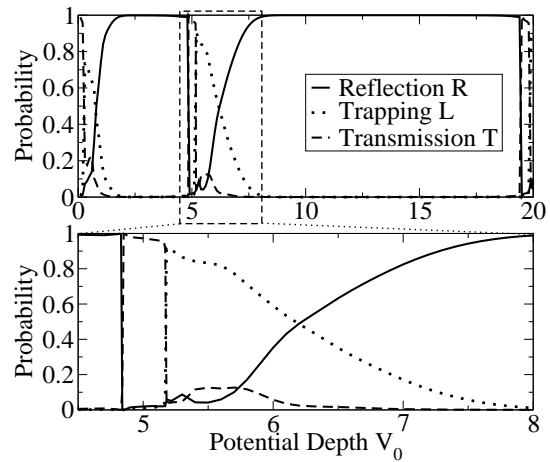}
\caption{\label{fig:rtlGPE} Reflection $R$, Transmission $T$ and Trapping $L$ ($L$ for '$L$'ocalized) as a function of $V_0$ and a fixed velocity $v=0.3$ by solving Eq.~\ref{GPE-soliton}. The lower picture zooms into one of the structures shown in the upper picture. The well width $2a=1$ is kept constant and we use $A=1$ and $g=-1$. The same parameters are used throughout the paper unless explicitly mentioned otherwise.}
\end{figure}

As physical observables we introduce the reflected ($R$), trapped/localized ($L$), and transmitted ($T$) fraction of the soliton, which are calculated at a time well after the initial impact of the soliton on the well (e.g. $t=166$):
\begin{eqnarray}
 R & = & \frac{1}{{\cal N}_s}  \int_{-\infty}^{-8a} dx |\psi(x,t)|^2 \nonumber \\
 L & = & \frac{1}{{\cal N}_s}  \int_{-8a}^{8a} dx |\psi(x,t)|^2 \nonumber \\
 T & = & \frac{1}{{\cal N}_s}  \int_{8a}^{\infty} dx |\psi(x,t)|^2  ,
\end{eqnarray}
with $R+L+T=1$.
Figure\ \ref{fig:rtlGPE} shows these quantities 
as a function of the depth of the well for a fixed initial velocity $v_{\mathrm{initial}}=0.3$. We consciously study the case of small velocity where $v_{\mathrm{initial}}^2/2\ll|\mu| = 0.5$. For the parameters of Ref.~\cite{2002Natur.417..150S} (see also Sec.~\ref{sec:level0}), this velocity amounts to $\approx 1.7${mm}$\,${s}$^{-1}$.

\begin{figure}
\includegraphics[width=0.7\columnwidth,angle=-90]{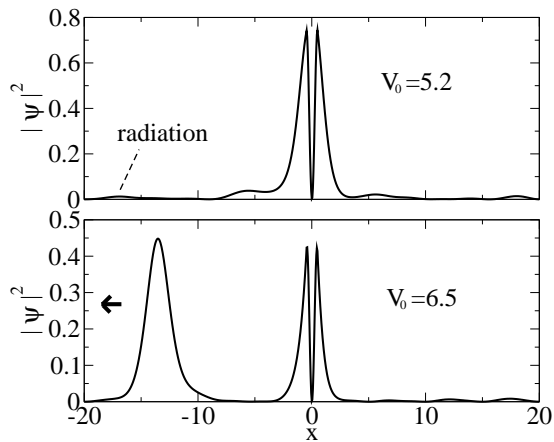}
\caption{The upper picture shows a snapshot of the condensate density at $t=77$ for $V_0=5.2$ where trapping is maximized in the second RL window. One can see the trapped mode in the first excited bound state (see text) and the radiation which stabilizes the trapped soliton. The lower picture shows partial trapping with a reflected soliton at t=65.}
\label{fig:radiation-reflected-soliton}
\end{figure}

The upper panel of Fig. \ref{fig:rtlGPE} shows several structures of similar form on a background of almost complete reflection. We thus call these structures reflection-trapping ($RL$) windows. In the following, we focus our discussion mostly on the second one as shown in the lower panel. For a certain range of $V_0$ the soliton reflects completely on the well. But by increasing the depth of the quantum well $R$ suddenly drops to zero while the transmitted fraction jumps to an absolute maximum. Further increase gives a sudden drop of $T$ to almost zero and most of the soliton is being trapped inside the quantum well. Then the trapping component $L$ starts to decrease while the reflected part increases. At least some of the reflected and transmitted amplitude in this part of Fig.~\ref{fig:rtlGPE} can be attributed to radiation, i.e.\ small amplitude waves. This becomes apparent in Fig.\ \ref{fig:radiation-reflected-soliton}, where snapshots of the density of the time-dependent wave function are shown. We will discuss the role that radiation plays in enabling trapping by carrying away kinetic energy in Sec.\ \ref{sec:level3}. At slightly larger $V_0$ we observe the co-existence of a reflected soliton with a trapped component together with radiation in the transmission channel becoming very small again.

Figure\ \ref{fig:radiation-reflected-soliton} reveals another remarkable feature: The condensate density has a single node localized close to the center of the well. Our simulations show that the number of nodes located in the well is a characteristic of each $RL$ window. Indeed we find that $RL$ windows appear around a critical well depth, where a linear bound state with the appropriate number of nodes is formed. In the first $RL$ window, the density reveals no node, the second one shows one, the third one shows two nodes and so on. The density of the soliton while located over the well is similar to the density functions for bound states of the Schr\"odinger equation in a quantum well. A more detailed analysis of the relation of the $RL$ window to linear resonances and nonlinear bound states of the well is given  further on in  this section.

\begin{figure}
\includegraphics[width=1.0\columnwidth,angle=0]{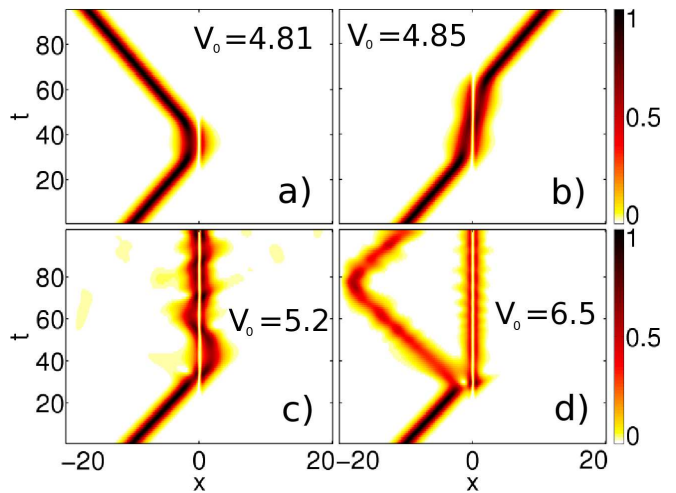}
\caption{\label{fig:carpet-var-V0} Time and spatial dependence of the condensate density $|\psi(x,t)|^2$ in gray scale (normalized to a maximum amplitude of 1) for four different $V_0$ as in Fig. \ref{fig:rtlGPE} but with $l_{box}=40$ The case of full reflection is shown in panel a) while the one for full transmission is given in b). Furthermore c) presents a fully trapped soliton while in d) the case of partial trapping and reflection is shown (the additional reflection towards the end comes from the hard wall boundary conditions).}
\end{figure}

The time-dynamics of the soliton are summarized in the density plot in Fig. \ref{fig:carpet-var-V0}. The pictures show the four different scenarios of full reflection, full transmission, full trapping and partial trapping. On the lower left picture the density sloshes around the center but a closer look brings up that the radiation reduces the amplitude of this oscillation and therefore stabilizes the trapped soliton. Furthermore, the position of the dip in density remains almost stationary, varying by not more than 5\% of the potential width. The reason for this is that for our choice of parameters the energy differences between the bound states in the well are large compared to any energy scale of the incoming soliton. Hence only one of these states can be populated, in case of Fig. \ref{fig:carpet-var-V0} it is the first excited state. 

\begin{figure}
\includegraphics[width=0.75\columnwidth,angle=-90]{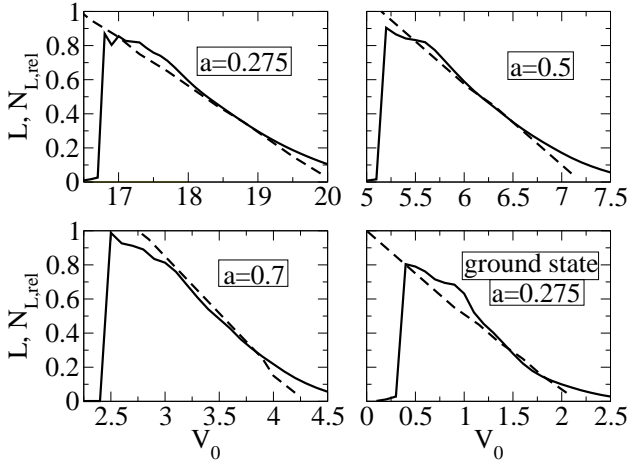}
\caption{\label{fig:compare-stationary} $L(V_0)$ (solid line) from the time-independent solutions and $N_{L,rel}(V_0)$ (dashed line) from the time-independent calculations for a fixed chemical potential $\mu=-\frac{1}{2}$. Both quantities show similar behavior, even for different potential widths and different bound states.}
\end{figure}

We now discuss the relation of the trapping phenomenon to (stationary) nonlinear bound states of the well.
Figure \ref{fig:compare-stationary} compares two different observables. The first one is the trapped component $L$ from the time-dependent simulations. The other one gives the relative number of particles $N_{L,rel}(V_0)$ in an eigenstate of the time-independent Gross-Pitaevskii equation for a fixed chemical potential $\mu$, that is set to the same value as the chemical potential of the free soliton
($\mu_{\mathrm{initial}} =-0.5$) in the time-dependent simulations. Specifically, $N_{L,rel}(V_0)$ is given by
\begin{equation}
 N_{L,rel}(V_0)=\frac{{\cal N}_E(V_0,\mu)}{{\cal N}_S(\mu)}
\end{equation}
where ${\cal N}_E(V_0,\mu)$ is the normalization constant (\ref{wf-normalization}) of the single-node stationary solution of Eq.~(\ref{GPE-soliton}) with the chemical potential $\mu$ while ${\cal N}_S(\mu)$ is the normalization of a free soliton with the same chemical potential.
In the numerical procedure $V_0$ is changed iteratively to keep the chemical potential at the desired value. The results for $N_{L,rel}(V_0)$ can then be compared with the relative number of trapped atoms $L$ we got from the time-dependent simulations. Even for different parameters the agreement between both graphs is very good.
These findings indicate that trapping is a resonant phenomenon with the chemical potential being the parameter of primary relevance.

\begin{figure}
\includegraphics[width=0.75\columnwidth,angle=-90]{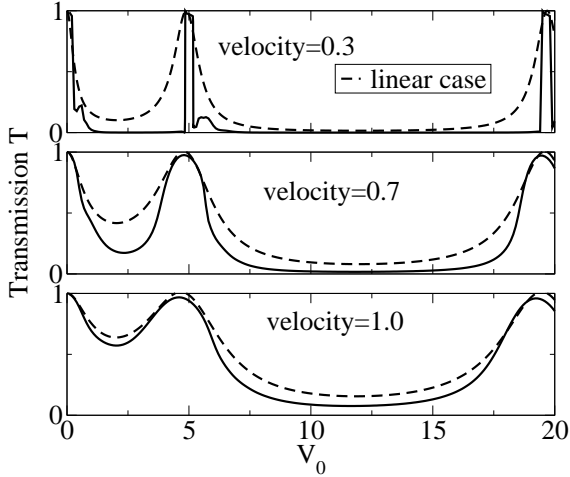}
\caption{\label{fig:transm-diff-V-nonlin_and_lin} Comparison of $T$ for solitons ($g=-1$, solid line) with the analytical solution  $T_{\mathrm{lin}}(V_0)$ [Eq.~(\ref{eq:linTransm})] for linear waves ($g=0$, dashed line). From top to bottom we increased $v_{\mathrm{initial}}$ of the incoming soliton. For increasing velocities the kinetic term in the Gross-Pitaevskii equation becomes dominant and therefore both curves approach each other.}
\end{figure}

Another feature in Fig. \ref{fig:rtlGPE} are the resonant transmission bands. They are closely related to the above-barrier transmission resonances in the linear Schr\"odinger equation, which is found from Eq. \ref{eq:1DGPE} for $g=0$. There one can find the analytical solution for the transmission \cite{schwabl:qm}
\begin{equation} \label{eq:linTransm}
 T_{\mathrm{lin}}(V_0)=\left[ 1+\frac{V_0^2}{v^2(v^2+2V_0)}\sin^2(2a\sqrt{v^2+2V_0})\right]^{-1} .
\end{equation}
In Fig.~\ref{fig:transm-diff-V-nonlin_and_lin} we compare the transmission for $g=0$ with the case of solitons at $g=-1$ at different velocities $v$ (see Fig. \ref{fig:transm-diff-V-nonlin_and_lin}). For very high velocities both curves approach each other. This is easily explained by the fact that the kinetic part in Eq. (\ref{GPE-soliton}) becomes much larger than the nonlinear term and therefore dominates the transmission spectrum. Thus decreasing $v$ increases smoothly the effects of the nonlinearity, in particular the formation of resonant transmission windows instead of transmission resonance lines. But their positions remain the same, which means that the nonlinearity just affects the shape of the transmission lines. We conclude that the basic mechanism of above-well shape resonances known from the linear Schr\"odinger equation remains valid for solitons.

\section{\label{sec:level2}Variational ansatz}

Goodman et al. \cite{2004PhyD..192..215G} studied soliton-defect interactions by simple two-mode models featuring a mobile soliton and a localized (trapped) mode. Here we extend this approach by including breathing of the trapped mode.

We approximate the well by an attractive delta potential, defined as
\begin{equation}
 V(x)=-\delta (x)V_0
\end{equation}
with $V_0>0$. With this simplification there is exactly one linear bound state for all potential depths.
Therefore we use an ansatz that splits the total wave function
\begin{equation}
 \psi=\psi_s +\psi_t.
\end{equation}
into a free soliton
\begin{equation}
 \psi_s = A_s \mathrm{sech} (A_s x-Q_s)e^{i\Phi_s}e^{iV_s x}
\end{equation}
and a trapped part
\begin{equation}
 \psi_t = A_t \mathrm{sech} (x/a_t)e^{i\Phi_t}e^{i\sigma_t \log\mathrm{cosh}(x/a_t)}
\end{equation}
that models a nonlinear mode that is localized at the well.
Here we introduced a particular form of chirping term $\log\mathrm{cosh}(x/a_t)$, which is capable of describing breathing modes. This can be used as a substitute for radiation effects which should allow the soliton to be trapped as it can transfer kinetic energy into another form of excitation. The choice for this particular form of the chirping term is consistent with Ref.~\cite{2004JPSJ...73..503S}.
This leads to our system's Lagrangian given by
\begin{widetext}
\begin{eqnarray}
\mathrm{L} & = & \int^\infty_{-\infty}dx\left\{ \frac{i}{2} \left( \psi^\dagger\frac{\partial}{\partial t}\psi -  \psi\frac{\partial}{\partial t}\psi^\dagger \right) -\frac{1}{2}\left|\frac{\partial}{\partial x}\psi\right|^2 + \frac{1}{2}\left|\psi\right|^4 - V(x)\left|\psi\right|^2   \right\} \nonumber \\
& = & -2A_t^2 a_t\dot{\Phi}_t -2A_t^2 a_t\dot{\sigma}_t (2-\log (4)) + 2A_t^2 \dot{a}_t \sigma_t -\frac{1}{3} \frac{A_t^2}{a_t} (1+\sigma_t^2) + \frac{2}{3} A_t^4 a_t \nonumber \\
& & - 2A_s\dot{\Phi}_s - 2\dot{V}_s Q_s +\frac{1}{3} A_s^3 - A_s V_s^2 \nonumber \\
& & + V_0\left\{ A_t^2 + A_s^2 \mathrm{sech}^2(Q_s) + 2A_t A_s \mathrm{sech}(Q_s)\cos (\Phi_s - \Phi_t)\right\}.
\end{eqnarray}
\end{widetext}
To obtain the equations of motion one has to solve the Euler-Lagrange equations
\begin{equation}
 \frac{d}{dt}\left( \frac{\partial \mathrm{L}}{\partial \dot{q}_i}\right)=\frac{\partial \mathrm{L}}{\partial q_i}
\end{equation}
for $q_i=A_s,\Phi_s,Q_s,V_s,A_t,\Phi_t,a_t,\sigma_t$.

This leads to
\begin{widetext}
\begin{eqnarray}
  \frac{d}{dt}A_s & = & V_0 A_s A_t \mathrm{sech}(Q_s)\sin(\Phi_s -\Phi_t )  \nonumber\\
 \frac{d}{dt}\Phi_s & = & \frac{1}{2}(A_s^2 -V_s^2)+V_0 \left[A_s\mathrm{sech}^2(Q_s) +A_t \mathrm{sech}(Q_s)\cos(\Phi_s -\Phi_t ) \right] \nonumber \\
\frac{d}{dt}Q_s  & = & A_s V_s \nonumber \\
\frac{d}{dt}V_s & = & -V_0 \left[A_s^2\mathrm{sech}^2(Q_s)\mathrm{tanh}(Q_s) +A_s A_t \mathrm{sech}(Q_s)\mathrm{tanh}(Q_s)\cos(\Phi_s -\Phi_t ) \right] \nonumber \\
\frac{d}{dt}a_t  & = & \frac{\sigma_t}{3a_t} + (2-\log (4)) \frac{V_0}{A_t}A_s\mathrm{sech}(Q_s)\sin (\Phi_s-\Phi_t) \nonumber \\
\frac{d}{dt}\sigma_t  & = & \frac{1}{3a_t^2}(1+\sigma^2) - \frac{A_t^2}{3} - \frac{V_0}{2a_t} +\frac{V_0}{2A_t a_t} A_s\mathrm{sech}(Q_s) \left[2\sigma_t\sin (\Phi_s-\Phi_t) - \cos (\Phi_s-\Phi_t)\right] \nonumber \\
\frac{d}{dt}\Phi_t  & = & -\dot{\sigma}_t (2-\log(4)) + \frac{\dot{a}_t \sigma_t}{a_t} - \frac{1}{6a_t^2}(1+\sigma_t^2) + \frac{2}{3}A_t^2 + \frac{V_0}{2A_t a_t}\left[ A_t + A_s\mathrm{sech}(Q_s) \sin (\Phi_s-\Phi_t) \right]. \label{Lagrangian-equations}
\end{eqnarray}
\end{widetext}

\begin{figure}
 \includegraphics[width=0.7\columnwidth,angle=-90]{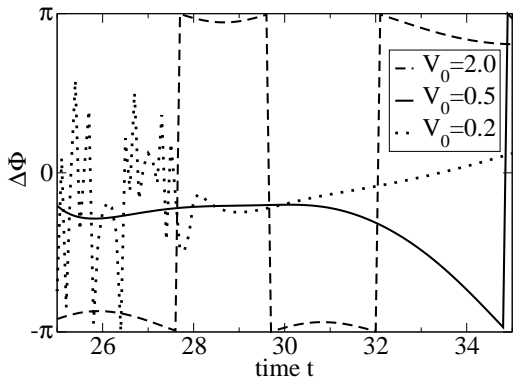}
\caption{The phase difference $\Delta\Phi$ for different potential depths and for the time when the incoming soliton reaches the quantum well at $t\approx 30$ from solutions of  Eqs.~(\ref{Lagrangian-equations}). The solid line and the dotted line show clearly a phase difference close to 0 where transmission respectively trapping occurs. An example for reflection with a turning point at $t\approx 29$ is given by the dashed line where $\Delta\Phi\approx\pm\pi$ (for initial conditions see text).}
\label{fig:phasediff}
\end{figure} 

Two of these eight dynamical variables can be eliminated due to conservation laws.
The amplitude $A_t$  can be found from the normalization of the wave function
\begin{equation}
{\cal N}_0=2A_s +2A_t^2 a_t ,
\end{equation}
since ${\cal N}_0$ is a constant of the motion. We obtain
\begin{equation}
A_t=\sqrt{\frac{{\cal N}_0}{2}-\frac{A_s}{a_t}} .
\end{equation}
Furthermore, it is not necessary to calculate $\Phi_t$ and $\Phi_s$ separately as the only interesting and physically important property is the phase difference 
\begin{equation} 
 \Delta\Phi=\Phi_t -\Phi_s .
\end{equation}
We note that the total phase $\Phi_t +\Phi_s$ has a trivial time dependence as it is canonically conjugate to the total energy (Hamiltonian), which is a further constant of the motion.
This way the problem left to solve consists of six coupled first-order ordinary differential equations.

\begin{figure}
 \includegraphics[width=0.9\columnwidth,angle=-90]{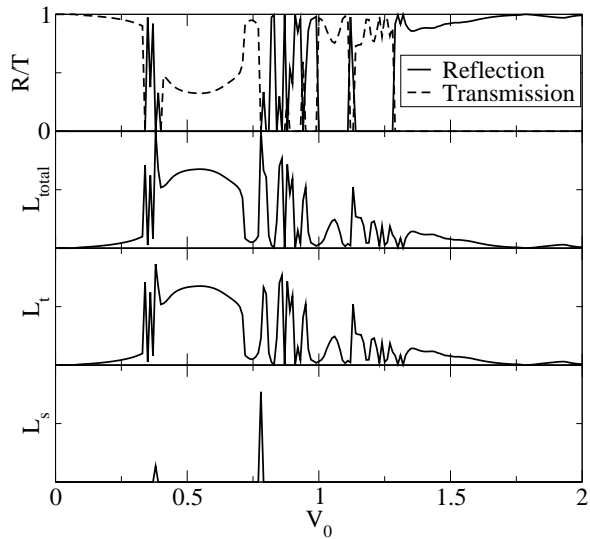}
\caption{Results for soliton scattering on a quantum well from the collective coordinate Eqs.~(\ref{Lagrangian-equations}). From top to bottom: Reflection R and transmission T vs $V_0$, the total trapping $L_{\mathrm{total}}$, fraction $L_t$ in the trapping mode, fraction $L_s$ in the soliton mode.}
\label{fig:different-L-and-rt}
\end{figure} 

\begin{figure}
 \includegraphics[width=1.0\columnwidth,angle=0]{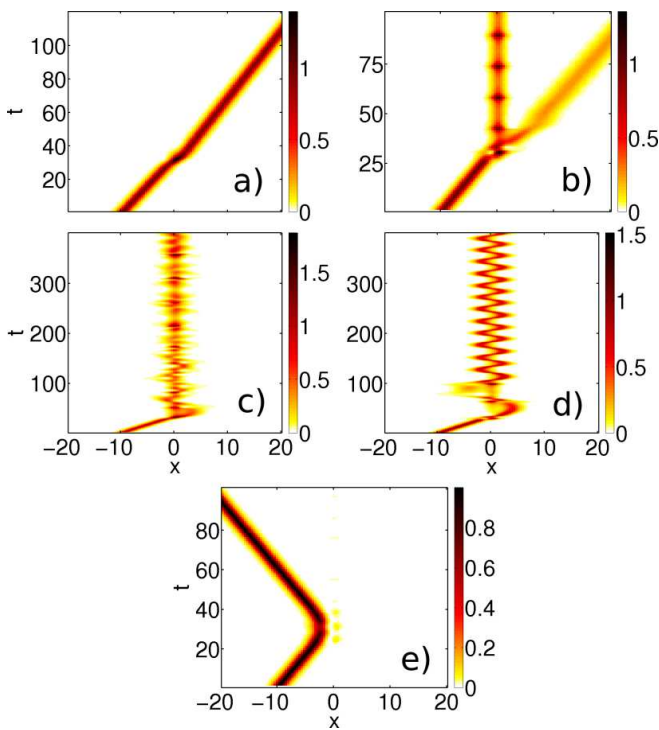}
\caption{(Color online) Condensate density as a function of time by solving  Eqs.~(\ref{Lagrangian-equations}), analogous to Fig. \ref{fig:carpet-var-V0}. For $V_0=0.2$ panel a) shows the typical situation a fully transmitted soliton. Partial trapping at $V_0=0.7$ is shown in b). Panels c) and d) present the cases for $V_0=0.38$ and $V_0=0.78$. The soliton is being trapped for c) due to the population of the trapped mode and sloshes around the well for d) while populating the soliton mode only. The last plot for $V_0=1.75$ shows the whole soliton being completely reflected.}
\label{fig:carpettwomode}
\end{figure}

This ansatz allows us to calculate the time dynamics of a soliton without solving the Gross-Pitaevskii equation directly. But, of course, this is still a very simple approximation and thus the results are not expected to be as accurate as the GP results. However they can give further insight in the mechanism involved.
Like in the previous section the initial velocity is set to $v_{\mathrm{initial}}=0.3$ and the initial position of the soliton is $Q_s=-10$.
We choose physically reasonable, small initial values for the parameters of the trapped mode ($A_t=10^{-4}$ and $a_t=10^{-2}$ for the simulations) in order to avoid numerical divergences. Furthermore we use $\Delta \Phi=0$ at $t=0$ but we find that the results do not depend on this initial choice.

Figure \ref{fig:phasediff} shows the phase difference $\Delta\Phi$ for different $V_0$ during the scattering process. For $V_0=0.2$ the soliton is being transmitted and we find $\Delta\Phi\ll 1$. Increasing the potential depth to $V_0=0.5$ results in (partial) trapping and $\Delta\Phi\ll 1$ while for a even larger $V_0=2$ there is full reflection ($\Delta\Phi\approx \pm \pi$). From Eq. (\ref{Lagrangian-equations}) we can see that the time dependence of the velocity $V_s$ highly depends on $\cos (\Delta\Phi)$. There we find that for a small phase difference the velocity does not change sign for all times and therefore the soliton transmits through the well or gets trapped. In the other case of $\Delta\Phi\approx \pm \pi$ the soliton can reflect from the well as the sign of the velocity can change. The difference between trapping and transmission however lies in the potential depth that determines how fast the trapped mode can be populated, i.e. large values for $V_0$ result in a faster population as can been seen in Eq. (\ref{Lagrangian-equations}) for $\dot{A_s}$ and $\dot{a_t}$. Therefore we can find a band between the reflection and the transmission regime where trapping can occur.

It is a well-know feature in collisions between bright solitons that a $\pi$-phase difference induces repulsion \cite{gordon:596}. This mechanism for reflection was discussed in Ref.\ \cite{2006EL.....73..321L}. In particular it avoids trapping.
Conversely, a resonant process with small phase difference is responsible for the population of the trapped mode.
This is consistent with the findings of the previous section where trapping was described as a resonant process.

To complete the comparison with the previous section, Fig.\ \ref{fig:different-L-and-rt} shows the reflection $R$, trapping $L$ and transmission $T$ as a function of the potential depth $V_0$. We see similar features as in Fig~\ref{fig:rtlGPE}. For very small $V_0$ the soliton is being transmitted almost completely (Fig.\ \ref{fig:carpettwomode}a) while for large $V_0$ full reflection (Fig.\ \ref{fig:carpettwomode}e) can be observed. Between both of these regimes we find a more complicated and interesting behavior. There, (almost) all the time one observes partial trapping of the soliton at the end of the simulation. Furthermore, we find two forms of trapping. The first case is the normal one. There we see that the trapping mode is being populated by the incoming soliton. The other fraction that remains in the soliton mode is moving either to positive or negative infinity (see Fig. \ref{fig:carpettwomode}b and \ref{fig:carpettwomode}c). In addition, another kind of trapping can be observed. In this situation the soliton mode oscillates around the delta potential (Fig. \ref{fig:carpettwomode}d). According to the numerical simulations this is the only event when full trapping occurs. 

We conclude that the basic ideas from the previous section are still valid: For small $V_0$ there is full transmission, then (partial) trapping and for very large $V_0$ the soliton reflects completely.

\section{\label{sec:level3}The trapping process}

In order to study the role of energy conservation and radiation in the trapping process,
we consider 
the energy functional
\begin{equation}
 E[\psi(x)]=\int dx \left[\frac{1}{2}\left|\frac{\partial}{\partial x}\psi(x)\right|^2 + V(x)|\psi(x)|^2 + \frac{g}{2}|\psi(x)|^4 \right]. \label{eq:tot_energy_general}
\end{equation}
We split this energy into different energy terms
\begin{equation}
 E[\psi_s]=E_{kin}^d+E_{kin}^v+E_{int}.
\end{equation}
These are defined as
\begin{eqnarray}
 E_{kin}^d & \equiv & \int dx \left[\frac{1}{2}\left|\frac{\partial|\psi(x)|}{\partial x}\right|^2\right] \nonumber\\
 E_{kin}^v & \equiv & \int dx \left[\frac{1}{2}\left||\psi(x)|\frac{\partial}{\partial x}\exp(i\theta(x,t))\right|^2\right] \nonumber\\
 E_{int} & \equiv & \int dx \left[-\frac{1}{2}|\psi(x)|^4 \right]
\end{eqnarray}
with $\psi(x)=|\psi(x)|\exp(i\theta(x,t))$.
The first term gives the contribution to the kinetic energy from the density variations while the second term represents a contribution from the phase gradient, which is connected to the superfluid velocity \cite{pitaevskiui2003bec}.
$E_{int}$ is the interaction energy.
Specifically for the soliton solution Eq.~(\ref{soliton_solution}) we find
\begin{equation}
 E_{kin}^d = \frac{1}{3}\sqrt{-g}A^3,\; E_{kin}^v = \frac{A}{\sqrt{-g}}v^2,\; E_{int} = -\frac{2}{3}\sqrt{-g}A^3
\end{equation}
and for the total energy
\begin{equation}
 E[\psi_s]=-\frac{\sqrt{-g}}{3}A^3+\frac{A}{\sqrt{-g}}v^2. \label{eq:sol_tot_energy}
\end{equation}

In particular, we find the universal ratio
\begin{equation}
 \frac{E_{int}}{E_{kin}^d} = -2 \label{eq:cond_sol}
\end{equation}
for the soliton solution.

\begin{figure}
 \includegraphics[angle=-90,width=0.9\columnwidth]{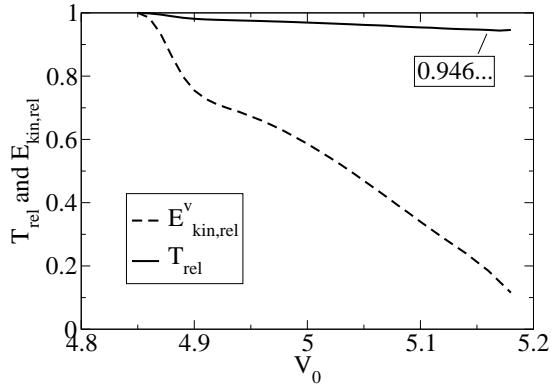}
\caption{Transmission $T_{\rm rel}(V_0)=T(V_0)/T(4.85)$ and velocity contribution to the kinetic energy $E_{\rm kin,rel}^v(V_0)=E_{\rm kin}^v(V_0)/E_{\rm kin}^v(4.85)$ relative to the values at $V_0=4.85$ where transmission is maximal. As the transmitted fraction of the incoming soliton decreases for deeper wells, so does its velocity $v_t$.}
\label{fig:rel_T_Ekin}
\end{figure}

\begin{figure}
 \includegraphics[angle=-90,width=0.9\columnwidth]{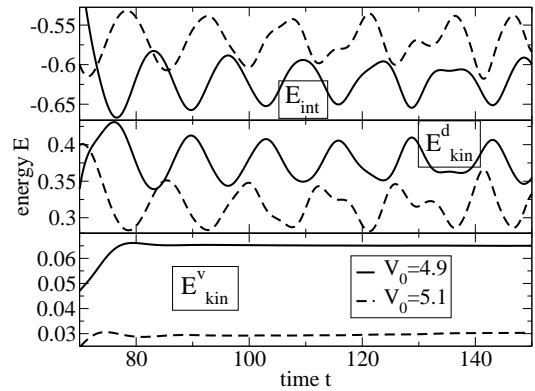}
\caption{Time dynamics of different kinetic and interaction energies for the transmitted part of the soliton for $V_0=4.9$ (solid line) and $V_0=5.1$ (dashed line) well after the collision ($t\approx 45)$). The soliton undergoes breathing oscillations after the scattering process that shown up as oscillations in the energy. For deeper wells $E^v_{kin}$ becomes smaller, i.e. the velocity of the transmitted soliton $v_t$ is smaller.}
\label{fig:wave_right}
\end{figure}
 \begin{figure}
 \includegraphics[angle=-90,width=0.9\columnwidth]{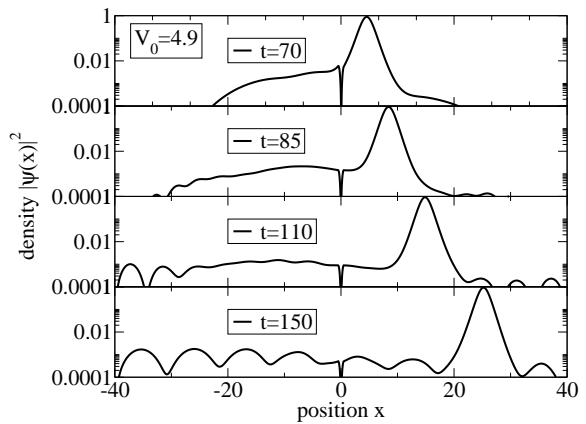}
\caption{Logarithmic density plot of the condensate for different times at $V_0=4.9$. A reflected component (radiation) is clearly visible.}
\label{fig:wave_function_V=4.9}
\end{figure}
 \begin{figure}
 \includegraphics[angle=-90,width=0.9\columnwidth]{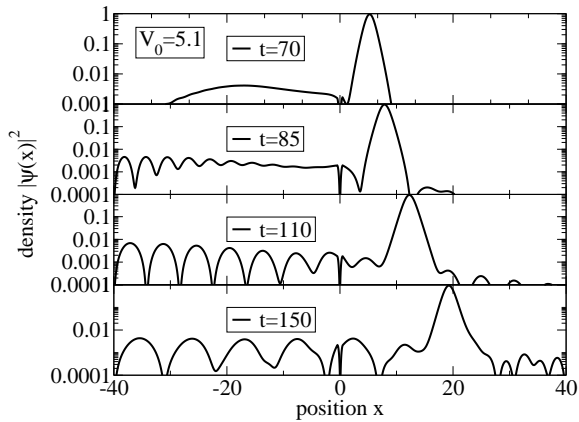}
\caption{Logarithmic density plot of the condensate for different times at $V_0=5.1$. Compared to Fig. \ref{fig:wave_function_V=4.9} the reflected part clearly larger.}
\label{fig:wave_function_V=5.1}
\end{figure}
\begin{figure}
 \includegraphics[angle=-90,width=0.9\columnwidth]{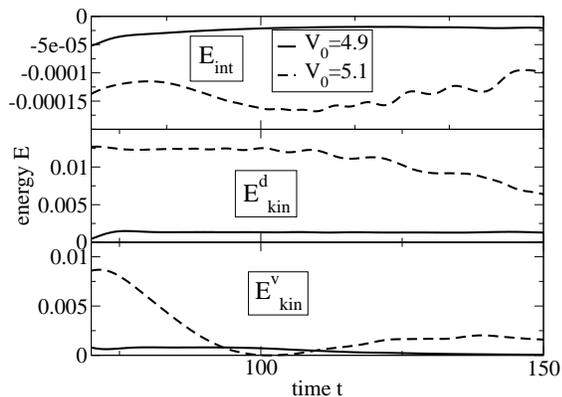}
\caption{Time dynamics of different kinetic and interaction energies for the reflected part of the soliton for $V_0=4.9$ and $V_0=5.1$. The ratio $\left|{E_{\rm int}}/{E_{\rm kin}^d}\right| \ll 1$, unlike what one expects from a soliton. As the interaction energy is small the reflected part is mainly radiation. This is only valid in the transmission window the reflected fraction is a soliton again for the case of trapping} 
\label{fig:wave_left}
\end{figure}
\begin{figure}
 \includegraphics[angle=-90,width=0.9\columnwidth]{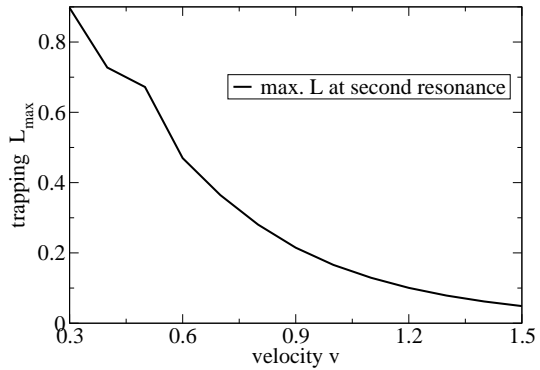}
\caption{The plot shows the maximum amount of atoms $L_{\rm max}$ that can be trapped around the second transmission resonance. $L_{\rm max}$ drops significantly with increasing velocity as proposed by Eq. (\ref{eq:energy_transm}).}
\label{fig:trapping_vs_velocity}
\end{figure}

We now show that radiation loss during a scattering event leads to a decreased velocity due to energy conservation. We consider a soliton (\ref{GPE-soliton}) with initial velocity $v_i$ that, during a collision event, suffers a small loss in amplitude due to radiation (small amplitude waves spreading away from the soliton).
The amplitude is reduced by the effect of radiation to $A_{t}=A-\epsilon$ with $0\ < \epsilon \ll A$. The energy of the transmitted soliton traveling with velocity $v_t$
is given by
\begin{eqnarray}
E_{t}[\psi] & = & -\frac{\sqrt{-g}}{3}(A-\epsilon)^3+\frac{(A-\epsilon)}{\sqrt{-g}}v_t^2 \nonumber\\
& = & \left[ -\frac{\sqrt{-g}}{3}A^3+\frac{A}{\sqrt{-g}}v_i^2 \right] +
\sqrt{-g} A^2 \epsilon \nonumber \\
& &  - \frac{-gA}{2v_i}\delta v+O(\epsilon^2)+O(\delta v^2), \label{eq:energy_transm}
\end{eqnarray}
where the last line has been linearized in $\epsilon$ and $\delta v \equiv v_t - v_i$. Identifying the term in square brackets as the energy of the initial soliton and assuming that radiation loss carries away a positive amount of energy (since the only negative contributions to energy could come from the nonlinear term, which is assumed to be small for radiation), we realize that the linear term in Eq.\ (\ref{eq:energy_transm}) must be negative. This leads to 
\begin{equation}
v_t \leq v_i - \frac{-gA}{2v_i}\epsilon < v_i  \label{eq:energy_transm2}
\end{equation}
since $\epsilon >0$. The slowing down of solitons after the collision can be seen in Fig.\ \ref{fig:rel_T_Ekin}, which compares the velocity part of the kinetic energy $E^v_{kin} \propto v_t^2$ and the transmission.

The same parameters are used as in Fig.\ \ref{fig:rtlGPE}, where the transmission window was found between $V_0=4.85$ and $V_0=5.2$. Fig.\ \ref{fig:rel_T_Ekin} shows that a small change in the transmission ($\approx 5\%$) results in a strong decrease of the transmitted soliton's velocity $v_t$ ($\approx 70\%$). 
Extrapolating Eq.\ (\ref{eq:energy_transm2}) beyond the regime of small $\epsilon$ we find  that for
\begin{equation}
 \epsilon \approx \frac{2v_i^2}{-gA} \label{eq:eps_crit}
\end{equation}
it predicts $v_t \approx 0$, which allows the soliton be trapped in the well.

For $v_i > \sqrt{\frac{-g}{2}}A$ we find that the right hand side of Eq.~(\ref{eq:energy_transm2}) is always positive as $\epsilon \leq A$. Therefore we expect that trapping is reduced until it vanishes for very high velocities $v_i$ when kinetic energy dominates over nonlinear energy contributions. Then the system becomes approximately linear and can be approximated by a single-particle.

In Figs \ref{fig:rel_T_Ekin}-\ref{fig:wave_left} we show results for the energy contributions after the soliton-well collision. In particular, Fig. \ref{fig:wave_right} gives energy contributions of the transmitted part of the soliton after the collision. There we can find that for both cases $\frac{E_{int}}{E_{kin}^d} \approx -2$. The curves show oscillations in energy which can be explained due to breathing of the soliton after the collision. In the bottom panel $E_{kin}^v$ is given for two $V_0$. We find again that the velocity $v_t$ decreases for larger $V_0$.

Snapshots of the condensate density for different times are given in Figs \ref{fig:wave_function_V=4.9} and \ref{fig:wave_function_V=5.1}. The incoming soliton transmits almost completely through the well, only a small portion is reflected as radiation. Furthermore both figures show once more that the radiation increases for deeper wells, i.e. the transmitted fraction is reduced.

Next we look at the reflected part in Fig. \ref{fig:wave_left}. There the ratio between interaction and the density contribution to the kinetic energy is
\begin{equation}
  \left|\frac{E_{int}}{E_{kin}^d}\right| \ll 1.
\end{equation}
Comparing to Eq. (\ref{eq:cond_sol}) this clearly indicates that the reflected part in this regime is not soliton-like. Instead the almost vanishing absolute value for the interaction term shows that the main contribution, the kinetic energy, is being carried by radiation as proposed in \cite{2006EL.....73..321L}.

The findings in this section help to understand the finite width of the transmission bands that we found in Sec. \ref{sec:level1}. If $v_i$ decreases the width of the transmission bands decrease as well, because less radiation is needed to trap the soliton.

\begin{figure}
\includegraphics[angle=-0,width=1.0\columnwidth]{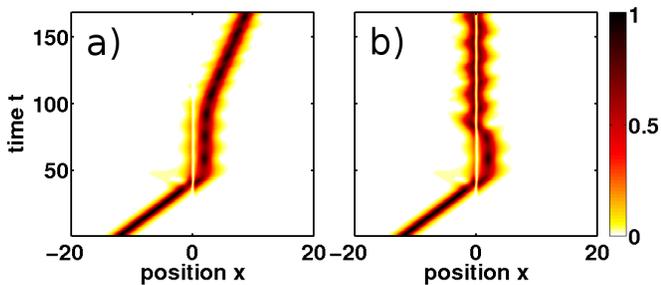}
\caption{Density plot for $V_0=5.182$ in a) and $V_0=5.183$ in b) ($l_{box}=40$). In part a) the soliton decelerates and remains at the edges of the well before it continues to move to the right hand side. In b) the soliton slows down and remains at the edges of the well before it decides to move back to get trapped by the well.}
\label{fig:carpet_transmitted_trapped}
\end{figure}

We illustrate the transmission and trapping behavior at the critical point for $V_0$ in Figure~ \ref{fig:carpet_transmitted_trapped}.
\newline

 \begin{figure}
 \includegraphics[angle=-90,width=1.0\columnwidth]{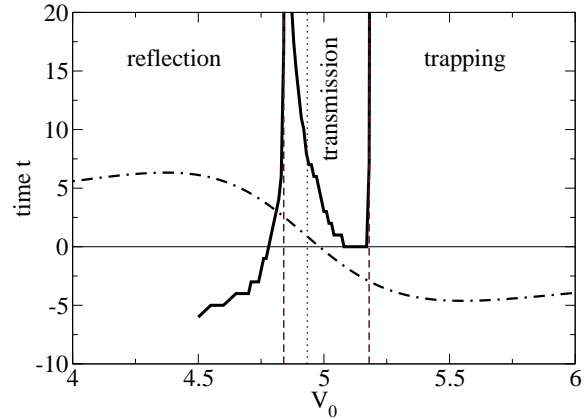}
\caption{Time delay of the soliton due to the existence of the potential well (solid line). These are results for the simulations in Sec. \ref{sec:level1}. Near the transition point from full reflection to full transmission (left vertical dashed line) the time delay increases very fast and becomes very large. This means the soliton remains in the vicinity of the trap on a very long time scale. Within the transmission region it decreases again until the trapping mechanism kicks in (right vertical dashed line). This picture for the nonlinear regime differs from the analytically calculated time delay for the linear case (dashed-dotted line) not only in the position and value of its maximum but also for the nonexistent negative time delay (right of the dotted vertical line).}
\label{fig:time_delay}
\end{figure}
\begin{figure}
\includegraphics[angle=-0,width=0.9\columnwidth]{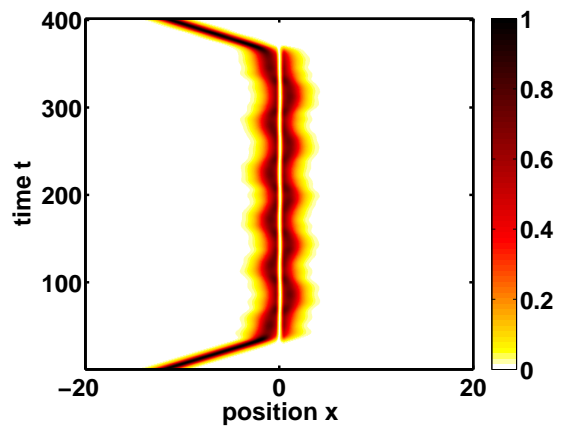}
\caption{Density plot for $V_0=4.842$ ($l_{box}=40$). The soliton decelerates and more than 99\% of the initial soliton is being trapped for a long time period (see also Fig. \ref{fig:time_delay}). This is connected to temporal trapping of a linear wave packet.}
\label{fig:delay_trapping}
\end{figure}

In addition there is a second mechanism to trap a soliton that is similar to the temporary trapping of a linear wave packet, which occurs at the boundary between reflection and transmission regions with radiation playing no role. For this situation we find that the soliton remains in the well for some time $t_d$ until it reflects. We can measure this time delay $t_d$ as the time the center of mass reaches the center of the well for the first time until it leaves the center again. Furthermore we see that by carefully adjusting the potential depth $t_d$ can be large enough to observe a temporally trapped soliton in experiments. Fig. \ref{fig:time_delay} shows the delay of the soliton during the transition through the well. It is remarkable that the whole soliton can be trapped with negligible losses due to radiation ($>99\%$). The losses are indeed much smaller than for the first trapping mechanism described before.
The time evolution of the density given in Fig. \ref{fig:delay_trapping} shows an example of the temporal trapping of the soliton. This delay within the well is analogues to the interaction free case for a traveling wave packet with velocity $v$ towards a well. There, an analytical expression for the time delay is known \cite{schwabl:qm} as
\begin{align}
 & t_d^{lin} = \nonumber \\
 & \left. \frac{\partial}{\partial E}\left[ \arctan \left(\frac{1}{2}\frac{\sqrt{E}}{\sqrt{E+V_0}} \tan\left(\sqrt{2(E+V_0}\right)\right)\right] \right|_{E=v^2/2}\; , \label{eq:time_delay_lin}
\end{align}
which is shown in Fig.\ \ref{fig:time_delay} as a dashed line.
This delay can be explained as a temporary trapping of the linear wave packet during which it oscillates between both ends of the well before it escapes again. However, due to the nonlinearity the position and the value for the maximum time delay differs significantly from the linear case. Furthermore the time delay of Eq. (\ref{eq:time_delay_lin}) becomes negative, which happens if the quantum well is deep enough to turn a quasi-bound state into a bound state which is in contrast to the nonlinear case. One should however note that in the linear case about 42\% of the wave packet reflects at the point of maximal time delay. For smaller velocities this value seem to converge towards 50\%. In addition the maximum time delay for the linear case lies well below the one for the nonlinear case, where it seems to diverge at the critical value for $V_0$. Therefore the connection to the nonlinear case is still unclear and needs further investigations that go beyond the scope of this work. 

We want to remark that although this would be an elegant way for lossless trapping of a soliton, Fig. \ref{fig:time_delay} also shows that the width of the this delay is very narrow and therefore harder to realize and to observe experimentally. In a BEC experiment with a small enough number of atoms it may be expected that superposition states will occur in this region \cite{weiss:010403}. Hence the other trapping method is favorable when it comes to experiments, even though one has to take into account minor losses.

\section{\label{sec:level4}Probing energy levels}

Trapping of soliton amplitude is sensitive to bound states in the well. Data presented in the previous sections has already suggested that trapping results from a resonant interaction of the soliton with a stationary defect mode, where the relevant energy scale is the soliton's chemical potential. By exploiting this resonant relationship, we are able to extract the bound-state energy by analyzing soliton scattering data. 
We proceed by comparing the scaled particle number of nonlinear bound state solutions with the trapped component after scattering a soliton with the same chemical potential.

\begin{figure}
\includegraphics[angle=-90,width=0.9\columnwidth]{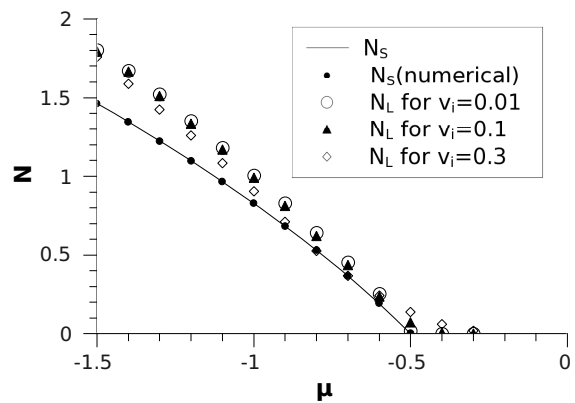}
\caption{The scaled number of trapped atoms $N_L$ vs.~$\mu_i$ from time-dependent simulations of Eq.~(\ref{GPE-soliton}) is compared with $N_S(\mu)$ for stationary solutions from Eq.~(\ref{eq:ns}). Stationary solutions were also found with our numerical code for checking numerical accuracy. Results taken over a range of initial velocities $v_i$ show a consistent picture. The most significant deviations occur at the onset of trapping around the location of the linear bound state at $E_b = -0.5$. This feature is most clearly distinguished for the smallest velocities.
}
\label{fig:probing-varv}
\end{figure}

In this section we model the defect as an attractive delta potential $V(x)= -V_0 \delta (x)$, which has only one bound state at $E_b=-V_0^2/2$. We solve Eq.\ (\ref{GPE-soliton}) with $g=-1$ with a soliton initial wave function (\ref{soliton_solution}), varying the amplitude $A$ and thus the chemical potential $\mu_i =g A^2/2$. After the soliton has scattered, we integrate the scaled particle number $N_L=\int_{-b}^b |\psi|^2 dx$ of the trapped component (choosing $b$ such as to capture at least 99\% of the initial soliton's normalization). In Fig.\ \ref{fig:probing-varv} we compare this data with the normalization $N_S = \int |\phi|^2 dx$ of a stationary localized solution $\psi(x,t)=\phi(x)\exp(i\mu_S t)$ of Eq.\ (\ref{GPE-soliton}) with the same chemical potential $\mu_S = \mu_i$.
We analytically find
\begin{equation} \label{eq:ns}
	N_{S}=2\left( \sqrt{-2\mu_S}-V_0 \right) ,
\end{equation}
which is shown as the full line in Fig.\ \ref{fig:probing-varv}.
The energy $E_b$ of the linear bound state [of Eq. (\ref{GPE-soliton}) with $g=0$] is found at the intersection of the line with the $\mu$ axis, i.e., $E_b = \mu_S$ at $N_S=0$. 

\begin{figure}
\includegraphics[angle=-90,width=0.9\columnwidth]{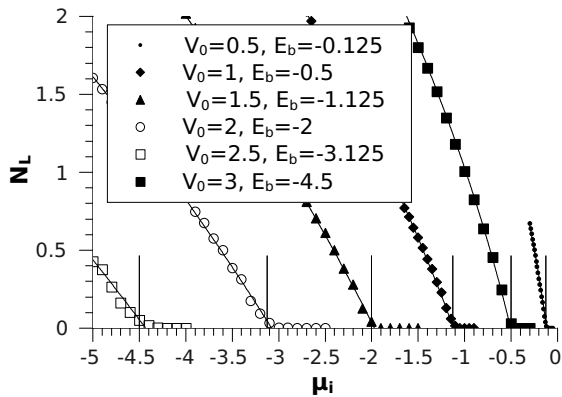}
\caption{Scaled particle number of the trapped component $N_L$ vs.\ $\mu_i$ after soliton-defect scattering as in Fig.~\ref{fig:probing-varv} for different values of the defect strength $V_0$. Short vertical lines indicate the energy of the bound state $E_b$ for each of the values of $V_0$. Square-root fits to the data (as explained in the text) provide estimates $\beta$ for $E_b$ from the scattering data. Values for $\beta$ found are -0.120,
-0.505, -1.12, -1.97, -3.08, -4.43, which should be compared with the corresponding exact values of $E_b$ given in the legend.}
\label{fig:probing-varV}
\end{figure}

As expected, trapping is observed in the time-dependent simulation only for $\mu_i \lessapprox E_b$ (Figs.~\ref{fig:probing-varv} and \ref{fig:probing-varV}) with the scaled particle number increasing with decreasing $\mu_i$, roughly following Eq.~\ref{eq:ns}. As seen in Fig.~\ref{fig:probing-varv} where data with a variety of different initial velocities is compared, the trapped component is systematically about 20\% larger than expected from the exact stationary solution. We have verified that the final state of the trapped component in the time-dependent simulations corresponds to a stationary solution with further reduced chemical potential compared to the initial $\mu_i$. While at this time we are not able to explain why this happens, the important result is that recording the trapped component as a function of $\mu_i$ allows us to locate the bound-state energy $E_b$.

In Fig.\ \ref{fig:probing-varV} we have plotted the trapped component as a function of the soliton's initial chemical potential $\mu_i$ for different trapping potentials. Least square fits of the data (data points with $N_L >0.003$ were included) to the functional form $N_{\mathrm{fit}}= \alpha \sqrt{\beta - \mu_i}$ provide estimates $\beta$ for the bound state energy $E_b$.

We expect that bound state energy levels of narrow potential wells of more general shape than the one studied here could be probed experimentally by scattering bright solitons using this scheme. For defects with more than a single linear bound state, we expect that only the least strongly bound one can be detected in this manner.

\section{\label{sec:level5}Conclusions}

In this work we have investigated the scattering of a bright soliton  on a linear defect in the context of matter-wave solitons. By numerical simulation and variational collective-coordinate studies, we have investigated the regime where the solitons are slow such that nonlinear energy scales dominate over kinetic energy and where the defect size is small compared to or of the same order as the soliton length. We have found a rich transmission -- reflection spectrum, which is strongly influenced by the level structure of the defect. In contrast to the scattering of linear waves, as in the scattering of single or independent atoms, part of the soliton can be trapped on the defect corresponding to the population of bound states or nonlinear localized modes.  

We find windows of transmission associated with above-well resonances in linear scattering. Nonlinear interactions modify the line shapes and lead to an abrupt onset of transmission. We have shown that particle loss of a few percent due to radiation leads to slowing down of the soliton due to energy conservation. A resonant coupling between the incoming soliton and bound states on the defect is identified as the mechanism leading to trapping and population of bound states. We have shown how this resonant coupling provides a way to experimentally probe bound states of an unknown localized potential well by scattering of nonlinear waves. 

In future work it will be interesting to study the subsequent scattering of two or more solitons or the scattering of solitons as we might expect additional effects of matter-wave enhancement or triggering the release of stored solitons. In addition, quantum many-particle effects like condensate fragmentation or macroscopic quantum tunneling could be investigated in the context of resonant soliton-defect scattering. While this question is very interesting it leads beyond the scope of the present work and thus is deferred to further study.

\begin{acknowledgments}
JB acknowledges stimulating discussions with Eric Heller, Andrew Martin, and Charles Adams. JB is supported by the Marsden Fund Council (contract MAU0706) from Government funding, administered by the Royal Society of New Zealand.
\end{acknowledgments}

\end{document}